# Comparación de las variables físicas que influyen en la rápida intensificación de los ciclones tropicales del Océano Pacífico nororiental durante el periodo 1970-2018

## Comparison of physical variables that influence the rapid intensification process of tropical cyclones over the Northeast Pacific Ocean during the 1970-2018 period

Mauricio López Reyes[1]  0000-0002-7698-0004
Ángel R. Meulenert Peña[2]  0000-0003-4504-8289

[1] Universidad Complutense de Madrid, Madrid, España. Departamento de Investigación del Instituto Frontera, Tijuana, México.

[2] Universidad de Guadalajara, Guadalajara, México.

**Resumen**

El estudio del proceso de rápida intensificación de los Ciclones Tropicales (CTs), fundamentalmente, es un tema actual y de poca investigación en México, donde intervienen factores térmicos y dinámicos a microescala y mesoescala. Lo anterior, debido a la poca mejora en los modelos de pronóstico de intensidad de los CTs y a la importancia de comprender y predecir su evolución, se emprendió una investigación de los CTs en aguas del océano Pacífico mexicano y sus factores que influyen en su proceso general de intensificación y rapidez, especialmente en la zona occidental que comprende los estados de Nayarit, Jalisco, Colima y al norte de Michoacán.

Los parámetros estudiados fueron; la temperatura superficial del mar (TSM), la profundidad de la capa de mezcla (PCM) y la cizalladura del viento (CV). A través de un proceso de selección de los CTs que cumplían la definición de Rápida Intensificación (RI) se procesaron y analizaron los parámetros antes mencionados, lo cual permitió identificar los valores umbrales que favorecen el proceso de RI. Se destacaron las anomalías de +7,6 % en TSM y +57 % en la PCM, si bien los valores de la cizalladura del viento no fueron determinantes, se observó una dominación en los factores térmicos TSM y PCM durante el proceso de RI.

Palabras clave: Rápida intensificación; Ciclones Tropicales; Pacífico Nororiental; Valores umbrales.

## Abstract

The study of the rapid intensification process of Tropical Cyclones (TCs) is a current, yet lacking research topic in Mexico, where thermal and dynamic factors at the microscale and mesoscale fundamentally intervene. Due to the little improvement in the TC intensity, forecasting models and the importance of understanding and predicting their evolution, an investigation was undertaken of the factors that influence the process of rapid intensification, and in general, the intensification of TCs in the Mexican Pacific Ocean, specifically in the western zone that includes states of Nayarit, Jalisco, Colima and north of Michoacán.

The parameters studied were, sea surface temperature (SST), mixing layer depth (MLD), as well as, wind shear (WS). Through a selection process of TCs that met the definition of Rapid Intensification (RI), the aforementioned parameters were processed and analyzed, which allowed identification of the threshold values that favor the rapid intensification process. The anomalies of + 7,6 % in SST and + 57 % in MLD stand out. Although the wind shear values were not low, a dominance in the thermal factors SST and MLD was observed during the RI process.

## 1. Introducción

Los Ciclones Tropicales (CTs) no son los más grandes ni más violentos sistemas meteorológicos que se generan en la atmósfera, ya que las bajas o ciclones extratropicales tienen una mayor extensión superficial, mientras que las Tormentas Eléctricas Locales Severas (TELS), que en ocasiones producen tornados, pueden ser mucho más intensas, acompañadas de otros fenómenos como fuertes ráfagas de viento y granizo de grandes dimensiones (Litta, 2012). Particularmente, cuando los CTs logran alcanzar la categoría de huracán mayor o por su definición en inglés «Major Hurricane» se pueden integrar las características de grandes dimensiones superficiales y las condiciones de severidad, por consiguiente, estos fenómenos son los más destructivos de los sistemas meteorológicos.

Dentro de las generalidades de los CTs, existen algunos que se intensifican de forma explosiva, este tipo de sistemas es definido por el National Hurricane Center (NHC) de los Estados Unidos de América, el cual señala lo siguiente:

Se define rápida intensificación (RI), a la condición meteorológica que sucede, cuando un ciclón tropical se intensifica dramáticamente en un corto periodo de tiempo, específicamente cuando el incremento de los vientos máximos sostenidos es de, al menos 30 nudos (55 km/h) en un periodo de 24 horas.

Este tipo de CTs son influenciados por diversos factores térmicos y dinámicos de la atmósfera y el océano, como la temperatura superficial del mar (TSM), cizalladura del viento (CV), humedad en la tropósfera (HT) y la disponibilidad de agua caliente en la capa de mezcla (DACCM).

En este trabajo se presenta un estudio sobre las variables oceánico-atmosféricas que permanecieron durante el proceso de RI de los CTs formados en el océano Pacífico nororiental frente a las costas de Jalisco, Nayarit, Colima y Michoacán, durante el período 1970 al 2018, y se compararon estos resultados con otro grupo de CTs que no presentaron RI, con el objetivo de obtener valores umbrales que detonen dicho proceso.

### 1.1. Estado del arte

Los efectos dañinos asociados a los CTs, particularmente de huracanes son: vientos fuertes, precipitaciones torrenciales, oleaje y marea de tormenta (Fossell, 2017).

Los vientos son una de las características que mejor identifican a los CTs, con excepción de los tornados (García et al., 2012), los huracanes son los fenómenos meteorológicos que presentan la mayor intensidad de viento, los que en ocasiones sobrepasan los 300 km/h (DeMaría et al., 2007), como lo fue el huracán Patricia en 2015, que alcanzó vientos sostenidos de 333 km/h.

Los fuertes vientos asociados a los huracanes ocasionan daños importantes debido a la gran fuerza que ejercen sobre cualquier estructura, por ejemplo, una placa cuadrada de 1 m² colocada ortogonalmente a un viento de 300 km/h estaría sometida a una fuerza media de 4170 N. La fuerza que ejerce el viento es proporcional al cuadrado de la magnitud de la velocidad (Sun et al., 2017), de ahí se deriva que aproximadamente bastaría un viento de 140 km/h para ocasionar una fuerza dos veces superior a la de un viento de 100 km/h, por lo tanto, la escala de peligrosidad de un huracán no aumenta de manera lineal.

El campo de viento tiene una estructura rotatoria alrededor de un centro denominado ojo del huracán que no cumple con leyes de simetría; es decir, el ojo del huracán no debe estar en el centro geométrico (Ávila, 2014). Los ciclones tropicales, sobre todo en sus etapas más débiles son bastante asimétricos en cuanto a la distribución de los campos de viento y las bandas de lluvia (García et al., 2017).

Los huracanes están íntimamente relacionados con los cuerpos de agua, especialmente los océanos, la intensidad y la extensión de la circulación del viento, que asociado a ellos provoca un fenómeno conocido como marejada de tormenta y oleaje que pueden afectar lugares muy lejanos de la zona de baja presión debido a la transferencia de energía en forma de ondas sobre la superficie del mar. La marea de tormenta es, para muchos especialistas, el efecto más destructivo asociado a los huracanes, ya que causa en promedio el 90% de las pérdidas de vidas humanas (CENAPRED, 2005).

Actualmente se han aprovechado las mejoras en los sistemas observacionales, lográndose una documentación más detallada. La realización de diferentes experimentos internacionales ha permitido conocer mucho más sobre estos sistemas meteorológicos, no obstante, el proceso de RI y de cambio de estructura de los CTs han tenido poco avance, especialmente en el océano Pacífico nororiental.

El estudio de los CTs de RI ha sido de especial interés en las agencias de investigación de los Estados Unidos de América en años recientes, en dichos estudio, J. Kaplan (2009) junto con M. DeMaria (2007) señalan que los factores oceánico-atmosféricos que estimulan el proceso de RI de los CTs estudiados en el océano Atlántico son principalmente, la Temperatura Superficial del Mar (TSM) misma que necesita tener valores por encima de los 26,5 °C (Knaff, 2009) aunado a una profunda capa de temperatura cálida que sirva como reservorio de energía para sostener la intensificación (Mainelli, 2008). Otro factor determinante para la rápida intensificación de un CT es la baja cizalladura del viento (Braun, 2013) ya que permite que la estructura interna del ciclón, especialmente en el radio de vientos máximos conserve la simetría y la convección profunda se fortalezca. Respecto a los estudios realizados en el Pacífico nororiental tropical, destaca la reciente investigación de (Oropeza y Raga, 2015) donde observaciones de altimetría satelital muestran que el Pacífico nororiental está poblado por remolinos oceánicos ciclónicos y anticiclónicos que afectan en la distribución espacial del calor oceánico e influye en la RI de los CTs.

Recientemente y con las investigaciones realizadas en los grandes huracanes Irma, María y José en 2017, se verificó lo predicho por (Braun, 2010) y (Braun, 2012) «una parte del núcleo de la tormenta debe volverse suficientemente húmeda antes de que se produzca una rápida intensificación», ya que en los eventos del 2017 se observó un desarrollo explosivo en la convección profunda en el anillo de vientos máximos antes del periodo de RI de estos CTs.

Los procesos de micro y mesoescala aún son complejos de modelar y por ello, los pronósticos de intensificación aún tienen gran incertidumbre; el problema radica en las múltiples variables que afectan la intensificación, la escala tan pequeña en la que estas ocurren y la dificultad de obtener datos suficientes y confiables para alimentar los modelos. El entender el mecanismo de intensificación de los CTs, implica desglosar la física de la microescala, por ejemplo, la liberación y transferencia de calor en las vecindades del ojo del huracán, los procesos que llevan a ciclos y entender los CTs, como sistemas dinámicos no lineales, en los que intervienen factores térmicos y dinámicos, lo que permitirá mejorar los pronósticos, principalmente los de intensidad, para salvaguardar vidas humanas y mitigar daños socioeconómicos.

Aparentemente, el cambio de intensidad de un huracán implica interacciones no lineales multiescala de diferentes fenómenos y variables (Marks, 1998; Wu, 2012). Estas interacciones incluyen a la TSM, el contenido de calor oceánico (CCO), como lo utilizó (Oropeza y Raga, 2015), en el Pacífico nororiental tropical, quien lo estimó en relación con la isoterma de 26,0 °C definida por (Leipper, 1972):

$$CCO = \rho_w C_w \int_{z=H_{26}}^{z} [T(z) - 26°C] dz \qquad (1)$$

Donde $\rho_w$ es la densidad media del agua del océano superior ($1026\ kg\ m^{-3}$), $C_w$ es el calor específico del agua salada a presión constante ($4178\ J\ kg^{-1} K^{-1}$), $T(z)$ es la temperatura del mar en función de la profundidad y los límites de integración corresponden a la temperatura a partir de la cual se induce la ciclogénesis y la altura de la superficie respectivamente. Aunque en este estudio no se utilizará la CCO, este parámetro está íntimamente relacionado con la TSM y la PCM cuyos datos si se incorporarán más tarde.

Otras de las variables que influyen en la RI y en general en la intensificación de los CTs es la CVV, la humedad ambiental, la dinámica y termodinámica del núcleo interno, la microfísica de las nubes y los procesos de interacción entre el aire y el agua oceánica (Chen, 2011) que no están bien representados en los modelos de huracanes, (Gall, 2013) menciona que incluso los modelos dinámicos actuales no alcanzan el nivel de habilidad de los modelos de intensidad estadística en muchos casos. Además, es notorio que, en los últimos años, los modelos de intensidad subestiman a los huracanes más fuertes (Chen, 2011). Otro de los fenómenos de mayor interés, tanto para los pronosticadores como para los investigadores, es la estructura interna del núcleo de la tormenta, como las bandas de lluvia, la variabilidad en el tamaño del ojo, especialmente los de poca dimensión, que se relacionan con los ciclos de remplazo de la pared del ojo, e influyen fuertemente en el desarrollo de la intensidad de los huracanes. (Abarca y Montgomery, 2014), encontró que los ciclos de emplazo de la pared del ojo se rigen en gran medida, por la dinámica de equilibrio axisimétrico de los anillos convectivos, además, en diversos artículos de Willoughby, Rozoff y Kepert, especialmente en (Kepert, 2010) se describió el mecanismo de contracción de la pared del ojo de la siguiente manera: «Las paredes del ojo, u otros anillos convectivos, se mueven hacia adentro como resultado del calentamiento adiabático diferencial entre el interior y el exterior» y a las leyes de conservación, por ejemplo, la del momento angular.

## 2. Materiales y Métodos

A continuación, se hablará de tres parámetros fundamentales que, con base en estudios previos, tienen influencia en el proceso de RI de los CTs; estos factores son los estudiados con detalle en el trabajo y se pueden clasificar debido a su carácter térmico y dinámico; TSM, PCM y CVV.

La TSM, es la temperatura en la superficie de océano, el significado exacto de superficie variará de acuerdo al método de medida usado, para el caso de este estudio se utilizarán las lecturas tomadas por satélite que miden la temperatura en el primer milímetro de la superficie de agua. Este proceso es de importancia primera para entender el proceso de intensificación de un CT y específicamente el de RI, ya que la TSM representa la fuente de energía disponible. DeMaria (2007) estableció que, para que la RI se lleve a cabo es necesario que la TSM sea mayor o igual a 27,5 °C, ya que el calor latente liberado es proporcional a la cantidad de agua que cambió de fase. Es importante mencionar que, en la región de estudio, la TSM está en constante variación, en cualquier escala temporal, debido a diversos factores como la batimetría, albedo, temporada anual y fenómenos como ENOS (EL NIÑO/Oscilación del Sur).

La capa de mezcla oceánica, es una capa en donde hay activa turbulencia que homogeniza algunos rangos de profundidades, esta región de mezcla se caracteriza por tener una temperatura prácticamente homogénea hasta la termoclina, lo que en teoría es un reservorio de calor que, traducido a la intensidad de los CTs, representa una fuente de energía que está en función de la profundidad, (Plata, 2016) llama a esta capa de mezcla como un recurso potencial para la intensificación progresiva y rápida de un huracán.

En condiciones típicas o medias históricas, la capa de mezcla en la región nororiental del Pacífico mexicano, especialmente en la región de estudio comprendida entre las latitudes 14 °N y 20 °N y a una distancia de 400 km de la costa, es de 30 m de profundidad, con una temperatura de 26,0 °C durante los meses entre agosto y octubre, según datos procesados en el European Centre for Medium Range Weather Forecasts (ECMWF) [disponibles en: https://www.ecmwf.int/en/research/climate-reanalysis].

La cizalladura del viento, es la diferencia en la velocidad del viento entre dos puntos en la atmósfera, típicamente se mide entre la altura de los 200 hPa y 850 hPa con la resta vectorial del viento. Para los objetivos del estudio, se analizan los resultados de la cizalladura del viento, es decir, el cambio en la magnitud de la velocidad del viento entre dos niveles de altura en términos del geopotencial.

La cizalladura del viento se define en la siguiente ecuación.

$$\vec{W}_s = \vec{V}_i - \vec{V}_j \qquad (2)$$

Donde $\vec{W}_s$ es el vector de cizalladura del viento, $\vec{V}_i$ es el vector velocidad del viento a la altura de 200 hPa y $V_j$ es el vector velocidad del viento a la altura de 850 hPa. De esta manera, se resta componente a componente se obtiene el vector de la cizalladura del viento con las componentes zonales y meridionales, $\hat{e}_u$ y $\hat{e}_v$ respectivamente en la ecuación (3)

$$\vec{W}_s = (u_i - u_j)\hat{e}_u + (u_i - u_j)\hat{e}_v \qquad (3)$$

En el estudio de la RI de los CTs, lo que interesa es conocer la norma de la cizalladura, por lo que la ecuación (4) representa la ecuación que se aplicará para determinarla.

$$\|\vec{W}_s\| = \sqrt{(u_i - u_j)^2 + (v_i - v_j)^2} \qquad (4)$$

La cizalladura del viento debe ser relativamente débil para que se favorezca el proceso de RI, es decir, no debe haber un cambio muy acentuado en la velocidad del viento con la altura, si la cizalladura del viento es mayor a 12,5 m/s, se trastoca la circulación del aire en torno a la zona de baja presión, un ciclón en formación se desorganiza y si ya está formado se debilita (Plata, 2006).

Con base en los elementos expuestos anteriormente, los objetivos del trabajo son, encontrar relaciones entre las variables térmicas y dinámicas de la atmósfera y la capa superficial del océano que estimulan el proceso de rápida intensificación, además de comparan los valores umbrales obtenidos para las variables de estudio de una muestra de CTs de RI, y otra muestra de ciclones que no fueron de RI en la misma región y espacio temporal.

Para realizar el estudio se definió una región del océano Pacífico nororiental, desde las costas mexicanas de los Estados de Nayarit, Colima, norte de Michoacán y Jalisco, con una distancia hacia el interior del Pacífico de 400 km, entre las coordenadas 16° y 23° latitud norte, y 110° y 100° longitud oeste, el área de estudio se observa en la figura 1. Esta es una región por la que, en cada Temporada Ciclónica, cruzan o se forman una buena cantidad de CTs, lo que permite trabajar con una muestra considerable.

Figura 1. Ubicación del área de estudio considerada.

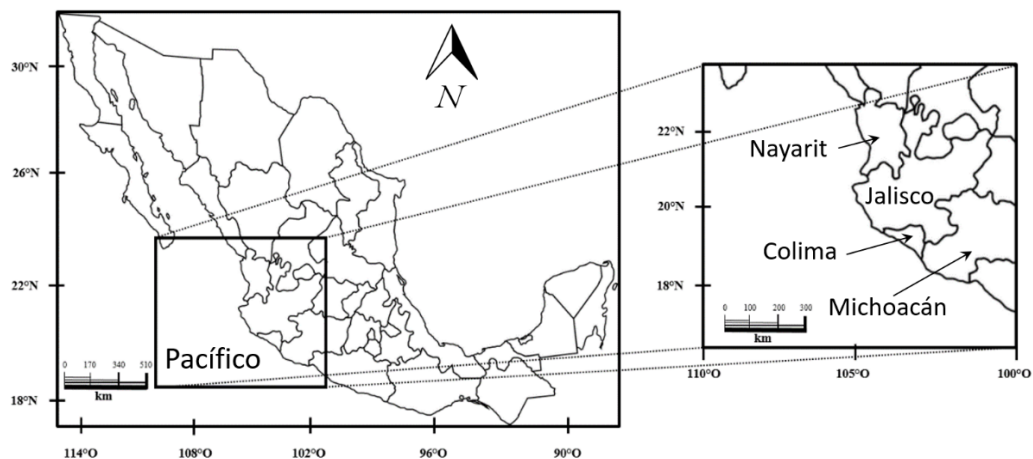

Se procesaron todos los CTs formados o que transitaron por la región de estudio durante el periodo de 1970-2018, con base en los archivos del NCH.

Los huracanes que cumplían con la definición de RI, fueron sometidos al análisis del entorno sinóptico en que se desarrollaron y la evolución en la rapidez de los vientos medios sostenidos, se utilizaron métodos numéricos de ajuste, que permiten determinar el periodo de tiempo en que alcanzan la RI. Con el método de mínimos cuadrados en lenguaje de Matlab, se obtiene la función de ajuste y el punto de inflexión, que indica la máxima rápida intensificación (MRI), de la variable rapidez del viento. Ver ecuación (5).

$$v(t) = \sum_{k=0}^{n} \alpha_k t^k, \quad \alpha_k \in \mathbb{R}, \ n = 4, 5 \tag{5}$$

$$a(t) = \frac{d}{dt}\sum_{k=0}^{n} \alpha_k t^k \tag{6}$$

Donde $\alpha_k$ son las constantes de ajuste y t es el tiempo parametrizado. Por cuestiones de procesamiento de datos y debido a que los datos obtenidos por el NHC de la velocidad del viento de los CTs se actualizan cada 6 horas, una hora parametrizada equivale a 6 horas en tiempo real.

La ecuación (6) representa la tasa de aceleración de los vientos y mediante un algoritmo programado en Matlab, se procesaron los datos de todos los CTs de RI para determinar el intervalo donde ocurrió la RI.

Se utilizó la base de datos «Reanalysis» de la (NOAA, 2017) [Datos disponibles en: https://psl.noaa.gov/data/gridded/data.ncep.reanalysis.html] para obtener los valores de las variables oceánicas y atmosféricas, específicamente mapas de contorno de TSM, PCM y cizalladura del viento, durante el periodo en que ocurrió la RI. Posteriormente, se calculan los valores medios de las variables para cada CT de RI y de una segunda muestra de 10 CTs que no presentaron periodo de RI. Los CTs de no RI se formaron en la región de estudio o su trayectoria la cruzó.

Finalmente, se procesan estadísticamente los parámetros encontrados, se utiliza la prueba de diferencia de medias con la distribución *t de student*, para muestras pequeñas, ver ecuación (7), y se aplica el criterio de prueba hipótesis para comparar los valores de las variables de los CTs RI y los no RI, ver las relaciones de la ecuación (8). Aunado a lo anterior, se clasifican los valores de las variables estudiadas con el algoritmo de clasificación K-means, para identificar los centros de los diagramas de dispersión y los vecinos más cercanos de cada clase, para comparar las variables uno a uno.

$$t = \frac{(\bar{x}_1 - \bar{x}_2) - \beta}{\sqrt{\frac{\sigma^2}{n_1} + \frac{\sigma^2}{n_2}}} \tag{7}$$

Donde $t$, es el estadístico de prueba, $\bar{x}_i$ son las medias de las muestras poblacionales, $\beta$ es el valor umbral propuesto, $\sigma^2$ es la varianza calculada por (7.1), y $n_i$ es el tamaño de las muestras.

$$\sigma^2 = \frac{(n_1-1)s_1^2 + (n_2-1)s_2^2}{n_1 + n_2 - 2} \tag{7.1}$$

Donde $s_i$ es la desviación estándar de cada muestra y al denominador $n_1 + n_2 - 2 = \omega$ se le conoce como los grados de libertad.

Para el caso de la prueba de hipótesis para 2 muestras, se utilizará la notación $H_0$ para la hipótesis nula y $H_1$ para la hipótesis alternativa, con un nivel de confianza de 95%.

$$H_0: \mu_1 - \mu_2 < \beta_i \\ H_1: \mu_1 - \mu_2 \geq \beta_i \quad (8)$$

Con base a la expresión (8) se desea demostrar que existe un valor umbral $\beta_i$, que diferencia los valores de las variables estudiadas de las muestras de los CTs de RI y no RI, para el caso de la TSM y la PCM, se mantiene el sentido de las desigualdades de (8) mientras que para la CVV se invierten las desigualdades.

En la figura 2 se observa el flujograma de la metodología utilizada en el trabajo.

Figura 2. Flujograma de la metodología utilizada.

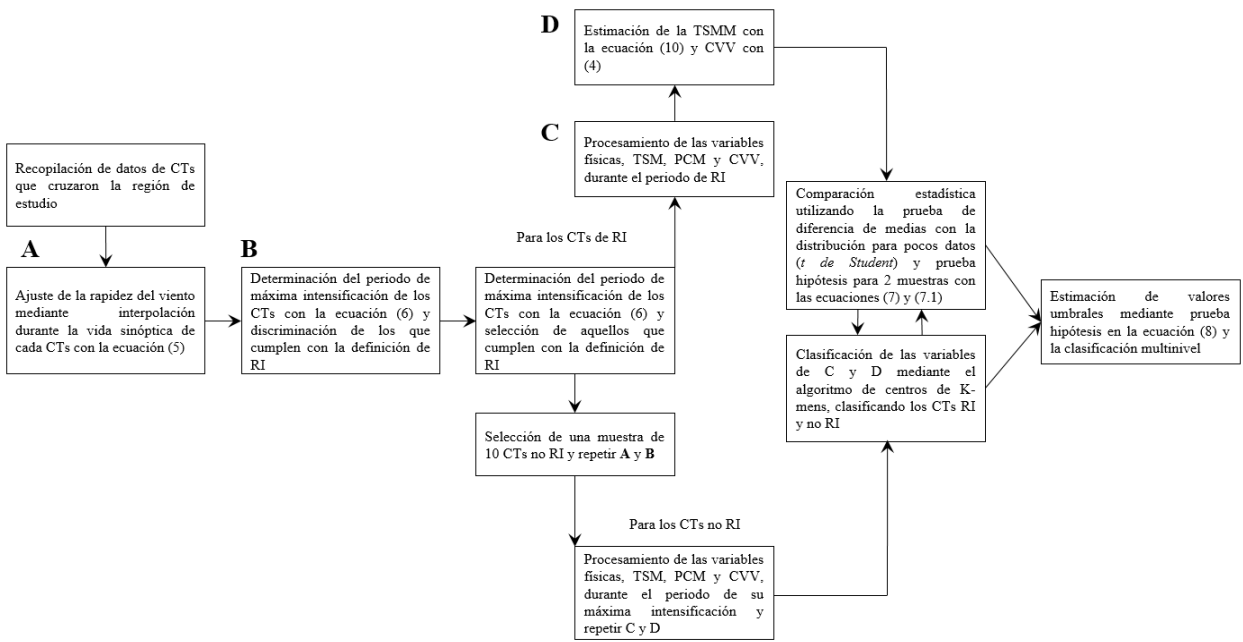

## 3. Resultados

### 3.1. *Caso de estudio del huracán Patricia*

Patricia se forma en aguas del Golfo de Tehuantepec a partir de una onda tropical y nace como CT el 20 de octubre del 2015. Durante los primeros días de vida, se mantuvo como tormenta tropical y comienza su periodo de RI el 22 de octubre en las coordenadas 101° 8´ O, 13° 40´ N.

Con base en los registros históricos del NHC, en la figura 3 a) se observa la evolución temporal de la rapidez del viento, así como la curva de ajuste $v(t)$ dada por la ecuación (9).

$$v(t) = -0.2t^4 + 2.0t^3 - 8.5t^2 + 1015 \quad (9)$$

Numéricamente se determinó que el instante de máxima intensificación de Patricia fue en el tiempo parametrizado[1] 9,2. En la figura 3 b) se señala con líneas verticales punteadas, el intervalo de RI en el cual, se estudiaron las variables físicas que intervienen en el proceso de RI.

Figura 3. a) Ajuste de la evolución de la rapidez del viento contra tiempo parametrizado durante el ciclo de vida del huracán Patricia. b) Pendientes en los extremos del intervalo de RI del huracán Patricia, con líneas punteadas se señala el periodo de RI en tiempo parametrizado.

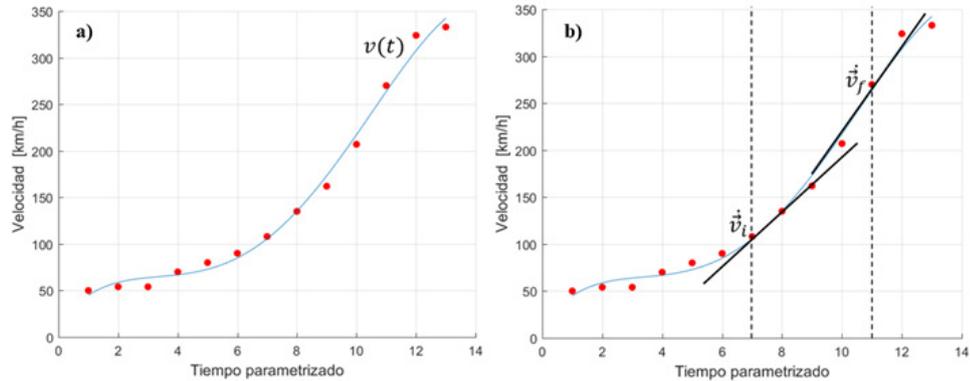

La figura 4 a) y b) corresponden a la TSM al inicio y final de la fase de RI respectivamente, se utilizó un método numérico discreto para determinar la Temperatura Superficial Media del Mar (TSMM), dicho método consiste en encerrar al CT en un cuadro de 300 km de lado con el ojo en el centro del cuadro, y tomar una muestra en cada nodo de una rejilla de 8 puntos muéstrales por lado, como se observa en la figura 5. La ecuación (10) muestra el proceso para obtener la TSMM en el cuadro descrito.

$$TSMM = \frac{1}{n^2} \sum_{i=1}^{n} \sum_{j=1}^{n} T_{ij} \qquad (10)$$

Donde *n* representa el número de puntos por lado de la malla.

Figura 4. a) Posición del CT Patricia al inicio de la fase de RI y TSM. b) Posición del CT Patricia al final de la fase de RI y TSM.

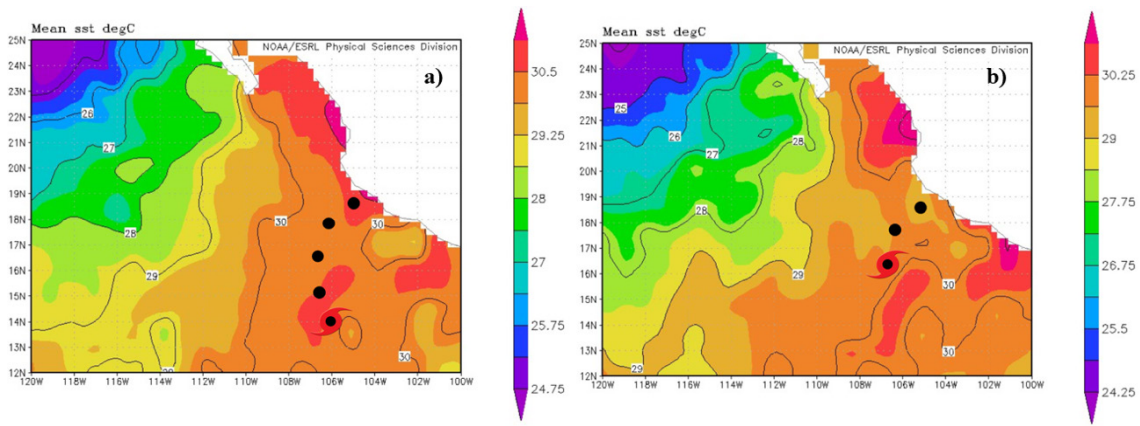

---

1. El tiempo parametrizado fue un concepto utilizado para facilitar el procesamiento de datos en Matlab. 1 hora parametrizada equivale a 6 horas reales.

Figura 5. Posición de Patricia durante el inicio de la fase de RI, TSM y rejilla de datos con centro en el ojo del huracán.

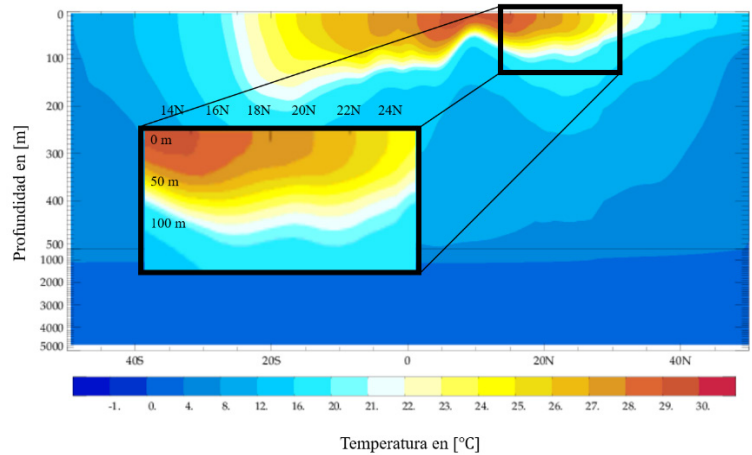

Fuente: Editada de los archivos del NHC.

Con base en los datos de TSM al inicio, durante y final de la fase de RI, se encontró que la TSMM fue de 31,2 °C, que representa una anomalía de +2,7 °C respecto a la media histórica de esa región.

En la figura 6, se presenta la temperatura y grosor de la capa de mezcla, se señala con un recuadro la región de interés, durante la intensificación de Patricia.

La latitud de estudio está comprendida entre los 14 °N y 20 °N, durante el intervalo de RI de Patricia, su coordenada meridional era de 17.0 °N, la temperatura de la capa de mezcla a los 40 metros de profundidad fue de 28,0 °C a 30,0 °C y a los 60 metros era de 26,0 °C a 27,0 °C. Los intervalos de temperatura se eligieron con base al criterio de la TSM mínima necesaria para estimular la intensificación de un CT, 26,0 °C (De Maria, 2007).

Figura 6. PCM en la región de interés durante el periodo de RI de Patricia.

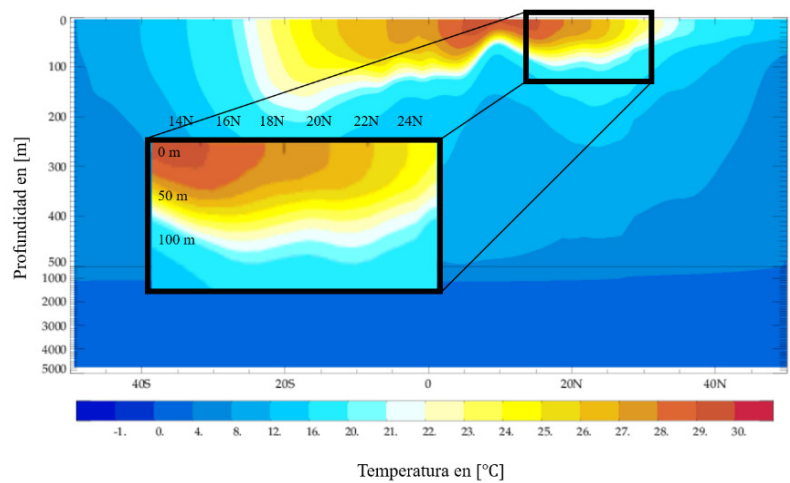

Al ubicar el centro del CT en el mapa, se delimitó una rejilla que incorpore un desplazamiento de un grado hacia cada dirección principal (Norte, Sur, Este y Oeste), se calcularon los componentes de la cizalladura del viento y su magnitud con base en las ecuaciones (3) y (4) respectivamente. En las figuras 7 y 8 se observan las curvas de nivel de las componentes zonales y meridionales del viento respectivamente.

Figura 7. Curvas de nivel de la componente zonal *u*, de la rapidez del viento en la región de estudio, sobrepuestas en la posición de Patricia. La imagen de la izquierda corresponde a la altura de 850 *hPa* y en la derecha la altura de 200 *hPa*

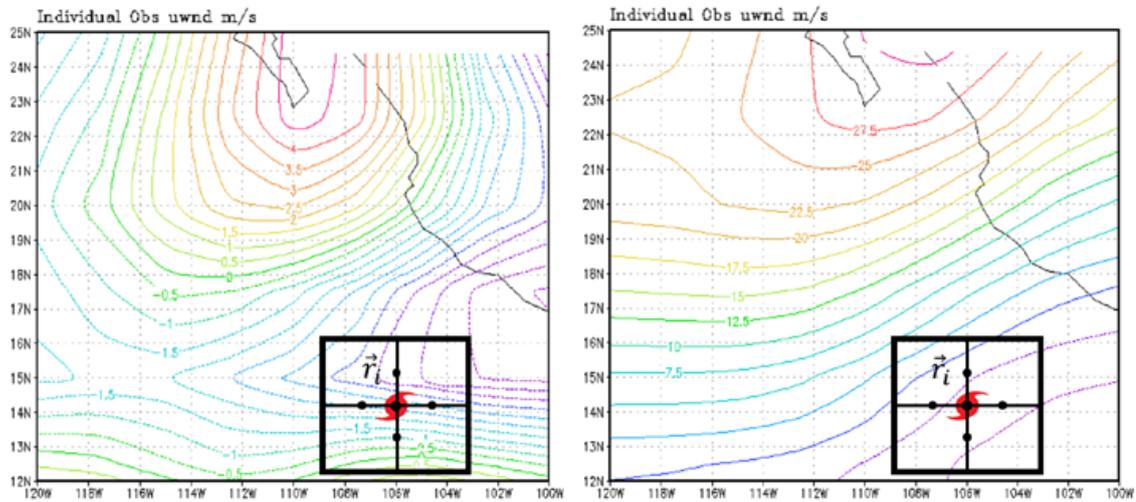

Figura 8. Curvas de nivel de la componente meridional *v*, de la rapidez del viento en la región de estudio, sobrepuestas en la posición de Patricia. La imagen de la izquierda corresponde a la altura de 850 *hPa* y en la derecha la altura de 200 *hPa*

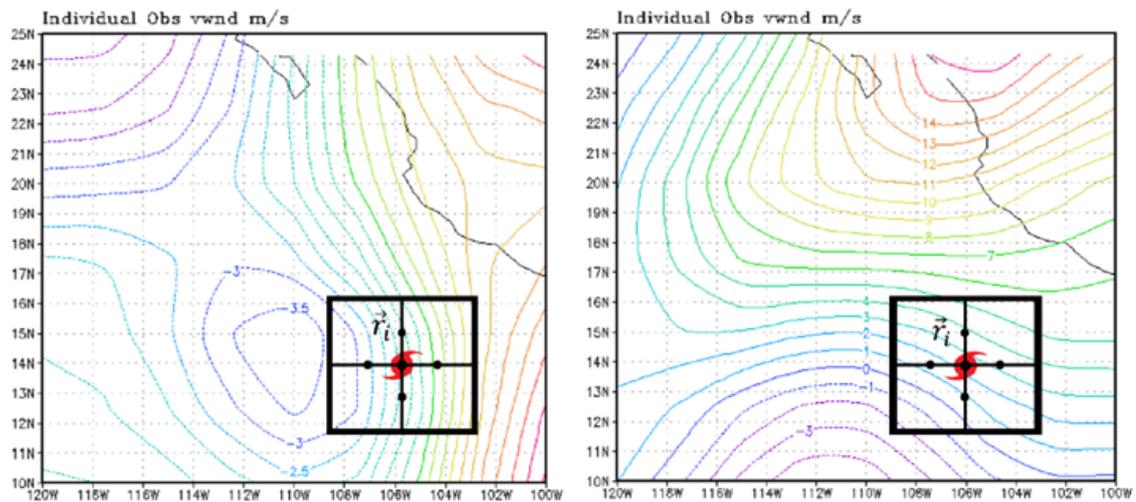

Se toma el valor absoluto de la diferencia de velocidades para obtener la magnitud de la cizalladura, por lo tanto, su magnitud media en una vecindad de un grado respecto al centro de Patricia fue de 8,8 $ms^{-1}$.

### 3.2. Comparativa entre los CTs de RI y los no RI

El proceso realizado en Patricia, se replicó en todos los CTs de RI y en un grupo muestra de CTs que no presentaron RI. En la tabla 1, se presenta el registro histórico de los huracanes RI que se estudiaron en el periodo de 1970-2018, así como, un grupo de CTs que no presentó RI en el mismo periodo.

Tabla 1. Ciclones tropicales RI y los que no presentaron RI registrados en el periodo de 1970-2018 en la región de estudio.

| Nombre del CT RI | Año y mes de formación | Nombre del CT no RI | Año y mes de formación |
|---|---|---|---|
| Patricia (Pa) | 2015-10 | Calvin (Ca) | 2014-08 |
| Odile (Od) | 2014-09 | Bud (Bu) | 2012-10 |
| Rick (Ri) | 2009-10 | Miriam (Mi) | 2012-09 |
| Kenna (Ke) | 2002-10 | Frank (Fr) | 2012-09 |
| Linda (Li) | 1997-09 | Andrés (An) | 2009-09 |
| Hilary (Hi) | 1997-09 | Fausto (Fa) | 2008-08 |
| Jova (Jo) | 1997-09 | John (Jo) | 2006-08 |
| Winifred (Wi) | 1992-10 | Olaf (Ol) | 2003-10 |
| Virgil (Vi) | 1992-10 | Marty (Ma) | 2003-09 |
| Kiko (Ki) | 1989-08 | Carlotta (Ca) | 2000-07 |
| Kiko (Ki*) | 1983-08 | | |
| Lorena (Lo) | 1983-09 | | |
| Tico (Ti) | 1983-10 | | |
| Norma (No) | 1981-10 | | |

Entre paréntesis, la abreviatura del nombre del CT. Fuente: NHC, Archive.

Una vez determinado el periodo de RI de cada CTs de la tabla 1, se calcularon las medidas de los factores térmicos y dinámicos. En la tabla 2 se presenta un concentrado referente a la TSMM en la región de estudio y la TSMM calculada durante el intervalo de RI de los CTs. Con base en información de la (NOAA, 2017), se estimó que la TSMM histórica en el área de estudio es de 28,5 °C en los meses de agosto, septiembre y octubre con lo cual se calcularon las anomalías.

Tabla 2. TSMM y anomalías respecto a la media temporal de la región de formación de los CTs con RI y sin RI.

| CTs con periodo de RI | | | CTs sin periodo de RI | | |
|---|---|---|---|---|---|
| Nombre del CT | TSMM [°C] | Anomalía de TSMM [°C] | Nombre del CT | TSMM [°C] | Anomalía de TSMM [°C] |
| Patricia (Pa) | 31,2 | +2,7 | Calvin (Ca) | 29,6 | +1,1 |
| Odile (Od) | 29,8 | +1,3 | Bud (Bu) | 28,9 | +0,4 |
| Rick (Ri) | 30,0 | +1,5 | Miriam (Mi) | 29,1 | +0,6 |
| Kenna (Ke) | 29,8 | +1,3 | Frank (Fr) | 27,4 | -0,9 |
| Linda (Li) | 30,1 | +1,6 | Andrés (An) | 28,2 | -0,3 |
| Hilary (Hi) | 28,9 | +0,4 | Fausto (Fa) | 28,5 | 0,0 |
| Jova (Jo) | 29,3 | +0,8 | John (Jo) | 27,9 | -0,6 |
| Winifred (Wi) | 28,7 | +0,2 | Olaf (Ol) | 27,5 | -1,0 |
| Virgil (Vi) | 29,7 | +1,2 | Marty (Ma) | 28,0 | -0,5 |
| Kiko (Ki) | 28,8 | +0,3 | Carlotta (Ca) | 27,3 | -1,2 |
| Kiko (Ki*) | 29,0 | +0,5 | | | |
| Lorena (Lo) | 28,3 | -0,2 | | | |
| Tico (Ti) | 29,0 | +0,5 | | | |
| Norma (No) | 27,8 | -0,7 | | | |

Fuente: NHC, Archive.

De los 14 CTs de RI estudiados, 13 de ellos se formaron en presencia del ENOS (El NIÑO/Oscilación del Sur), se destacan los episodios de los años 2015 y 1997 donde se formaron los poderosos huracanes Patricia y Linda, respectivamente. Al comparar los datos de la tabla 2, se observa que estos huracanes estuvieron influenciados por las mayores anomalías de la TSMM.

El promedio de la TSMM de los CTs de la tabla 2 fue de 29,4 °C que representa una anomalía media de +1,4 °C, esta anomalía es superior a la obtenida por (Montgomery, 2012) en la temporada ciclónica del 2011 en el Atlántico, cuyo valor fue de +0,7 °C. Respecto a la diferencia entre la TSMM entre los CTs de RI y los no RI se obtuvo, 1,2 °C de diferencia.

Respecto a la disponibilidad de agua cálida en la capa de mezcla, antes de la termoclina de 26,0 °C durante la formación de los CTs de RI, los datos se concentran en la tabla 3.

Tabla 3. Profundidad de la capa de mezcla para temperaturas de 26,0 °C o superiores.

| CTs con periodo de RI | | CTs sin periodo de RI | |
|---|---|---|---|
| **Nombre del CT** | **PCM [m]** | **Nombre del CT** | **PCM [m]** |
| Patricia (Pa) | 70 | Calvin (Ca) | 40 |
| Odile (Od) | 50 | Bud (Bu) | 40 |
| Rick (Ri) | 50 | Miriam (Mi) | 45 |
| Kenna (Ke) | 40 | Frank (Fr) | 35 |
| Linda (Li) | 65 | Andrés (An) | 45 |
| Hilary (Hi) | 50 | Fausto (Fa) | 35 |
| Jova (Jo) | 50 | John (Jo) | 35 |
| Winifred (Wi) | 50 | Olaf (Ol) | 30 |
| Virgil (Vi) | 50 | Marty (Ma) | 40 |
| Kiko (Ki) | 60 | Carlotta (Ca) | 35 |
| Kiko (Ki*) | 60 | | |
| Lorena (Lo) | 65 | | |
| Tico (Ti) | 65 | | |
| Norma (No) | S/D | | |

El CT Norma no tiene disponible el dato de PCM. Fuente: NHC, Archive.

El promedio de la PCM para temperatura mayores a 26,0 °C fue de 62 m, cabe resaltar que, en los episodios de Patricia y Linda, la termoclina de los 28,0°C tenía una profundidad de 50 m sobre la región de estudio, lo cual representa un gran reservorio de energía que alimentó su rápida y prolongada intensificación, en comparación con el estudio de (Montgomery, 2012) en el Atlántico, la PCM en los episodios de RI de la temporada 2011 fue de 55 m. En el Pacífico, no existen datos para compararse.

En la tabla 4 se muestran los valores de la magnitud de la cizalladura del viento durante el intervalo de RI, es importante destacar que algunos huracanes anteriores a Linda 1997, carecen de información sobre todo en las lecturas tomadas en los 200 hPa.

Tabla 4. Valores estimados de la cizalladura del viento.

| CTs con periodo de RI | | CTs sin periodo de RI | |
|---|---|---|---|
| **Nombre del CT** | **CVV [m/s]** | **Nombre del CT** | **CVV [m/s]** |
| Patricia (Pa) | 8,8 | Calvin (Ca) | 11,2 |
| Odile (Od) | 8,4 | Bud (Bu) | 10,2 |

| CTs con periodo de RI | | CTs sin periodo de RI | |
|---|---|---|---|
| **Nombre del CT** | **CVV [m/s]** | **Nombre del CT** | **CVV [m/s]** |
| Rick (Ri) | 9,1 | Miriam (Mi) | 10,9 |
| Kenna (Ke) | 10,2 | Frank (Fr) | 9,2 |
| Linda (Li) | 7,9 | Andrés (An) | 12,4 |
| Hilary (Hi) | 8,9 | Fausto (Fa) | 13,5 |
| Jova (Jo) | 11,0 | John (Jo) | 9,8 |
| Winifred (Wi) | 8,9 | Olaf (Ol) | 11,6 |
| Virgil (Vi) | 8,2 | Marty (Ma) | 10,5 |
| Kiko (Ki) | 6,8 | Carlotta (Ca) | 10,5 |
| Kiko (Ki*) | S/D | | |
| Lorena (Lo) | 8,5 | | |
| Tico (Ti) | 6,6 | | |
| Norma (No) | S/D | | |

El CT Norma no tiene disponible el dato de PCM. Fuente: NHC, Archive.

Con base en los datos reportados de la cizalladura del viento, se calculó que la media de todos los CTs de RI fue de 8,8 $ms^{-1}$. Este valor, se encuentra dentro del rango favorable para la intensificación de un CT. En comparación con el estudio de (Montgomery, 2012) en el Atlántico, la cizalladura del viento en los episodios de RI de la temporada 2011 fue de 6,8 $ms^{-1}$, lo que representa valores ligeramente menores que los obtenidos en este estudio.

Mediante la aplicación de la prueba de diferencia de medias, con el valor umbral propuesto para la prueba de hipótesis, $\boldsymbol{\beta_1 = 1,2°C}$, que representa la diferencia entre la TSMM para todos los CTs de RI estudiados, comparada con los no RI; tenemos que $\boldsymbol{H_{01}: TSMM_{RI} - TSMM_{no\ RI} < \beta_1}$ y $\boldsymbol{H_{11}: TSMM_{RI} - TSMM_{no\ RI} \geq \beta_1}$, al aplicar la ecuación (7), se demuestra que, $\boldsymbol{H_{11}}$ es verdadera con un nivel de confianza del 95%, por lo tanto podemos considerar que el valor umbral de la TSMM mínima, que favorece la RI en los CTs estudiados, debe ser superior a $\boldsymbol{1,2\ °C}$ respecto a las condiciones de los CTs no RI.

Para el caso de la PCM, el valor umbral propuesto es, $\boldsymbol{\beta_2 = 30\ m}$, que representa la diferencia media de la PCM durante los episodios de intensificación de los CTs de RI y los no RI, entonces las hipótesis son: $\boldsymbol{H_{02}: PCM_{RI} - PCM_{no\ RI} < \beta_2}$ y $\boldsymbol{H_{12}: PCM_{RI} - PCM_{no\ RI} \geq \beta_1}$, nuevamente se verifica la veracidad de la hipótesis alternativa $\boldsymbol{H_{12}}$, por lo tanto, el valor umbral de 30 metros «extras» de profundidad con una temperatura mayor a 26,0 °C, es significativo para la RI de los CTs estudiados, estos resultados son congruentes con lo mencionado por (Knaff et al., 2011) hay épocas del año donde los factores térmicos dominan sobre los dinámicos en la intensificación de los CTs. Un estudio complementario interesante, sería la comparación de los resultados de las variables térmicas de este trabajo, con lo encontrado por (Oropeza y Raga, 2015) el 74% de los CTs que alcanzaron la rápida intensificación, tuvieron interacción directa y/o indirecta con remolinos oceánicos anticiclónicos, y de ellos el 86% alcanzó la categoría de huracán mayor.

Las inferencias anteriores se refuerzan con los resultados del algoritmo de clasificación K-means, en la figura 9-A, al clasificar todos los CTs estudiados con base en los 2 centros principales respecto a la TSMM y la PCM, se obtiene que 86% de los CTs con características de RI, se encuentran más cerca del centro de los RI, es decir, 12 de los 14 CTs de RI, tienen los valores umbrales de TSMM y PCM más cercanos a la media de la clasificación de RI, por otro lado, el 90% de los CTs no RI, también se encuentra dentro de la clasificación de no RI, esto permite inferir que, para la muestra estudiada, los valores de la TSMM y PCM, para la termoclina de 26,0 °C, son bastante congruentes con la tasa de intensificación de los dos grupos de CTs.

En cuanto a la CVV, se obtuvo por comparación entre medias, el umbral de $\beta_3 = -2,5\ ms^{-1}$, por lo que las hipótesis tienen la forma: $H_{03}: CVV_{RI} - CVV_{no\ RI} > \beta_3$ y $H_{13}: CVV_{RI} - CVV_{no\ RI} \leq \beta_3$ al aplicar la ecuación (7) con una certidumbre del 95%, se demuestra que la hipótesis, $H_{13}$ es verdadera y en consecuencia el valor umbral, $-2,5\ ms^{-1}$ es significativo para la RI de los CTs. Si bien, valores altos de la cizalladura, rompen con la transferencia de calor interna cerca del núcleo de la tormenta (Kepert, 2010), un valor muy bajo o nulo de la cizalladura, inhibe la entrada de aire húmedo, estrangulando la intensificación del sistema tropical (Knaff et al., 2011), tal vez sea por ello que el valor umbral encontrado en este trabajo para la CVV, es bajo en comparación con los CTs no RI, pero aun permite el desarrollo de flujos de humedad y calor en el interior de la tormenta.

Con base en la figura 9-B y 9-C, se observa cómo los factores térmicos para los CTs de RI se ubican hacia los valores mayores del eje horizontal, para la PCM y TSMM respectivamente, mientras que en el eje vertical correspondiente a la CVV, los valores mayores de este parámetro lo tienen los CTs de no RI, al utilizar el algoritmo K-means para clasificar los CTs de los dos grupos, se observa de la figura 12-B que, 93% de los CTs de RI entran en la clasificación del centro de RI y 100% de los CTs no RI entran en la clasificación del centro de los no RI, análogamente para la figura 12-C, 93% de los CTs de RI entran en la clasificación del centro de RI y 70% de los CTs no RI entran en la clasificación del centro de los no RI.

Figura 9. Diagramas de dispersión entre las variables estudiadas para los grupos de CTs de RI y no RI. La circunferencia continua encierra el grupo de los CTs de RI, y la circunferencia punteada a los CTs no RI.

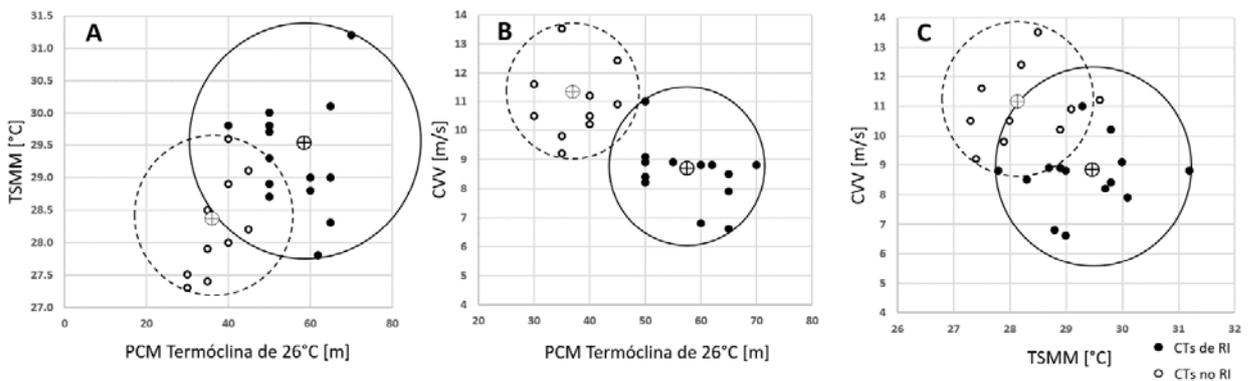

Fuente: Elaboración propia.

En la figura 10 se muestran los diagramas de cajas y bigotes para las variables estudiadas entre los CTs de RI y lo no RI, destaca en los casos térmicos, el sesgo de los valores más altos mostrados en los cuartiles $Q_1$, $Q_2$ y $Q_3$, tanto de la TSMM como de la PCM para los RI representados en a1) y a2), respecto a los no RI en b1) y b2). Por otro lado, en c1) y c2) se aprecia un considerable aumento en los valores de la CVV. para los CTs no RI, esto se cuantifica con la ubicación de los cuartiles $Q_2$, $Q_3$ y el rango intercuartil $R$, entre ambas muestras para los CTs. Con ayuda de los diagramas de la figura 10, específicamente en a1) y b1), se nota como la clasificación por el método K-means se vuelve menos evidente, pues hay varios CTs de RI y no RI, cuyos valores de la TSMM estaban en un rango compartido, por ejemplo, entre el cuartil $Q_1$ y $Q_2$ de a1) y el $Q_3$ y $Q_4$ de b1) ya que coinciden varios CTs de ambas categorías.

Figura 10. Distribución por cuartiles y valores extremos de la TSMM, PCM para la termoclina de 26,0 °C y CVV para los CTs de RI a1), a2) y a3) y par los CTs no RI b1), b2) y b3).

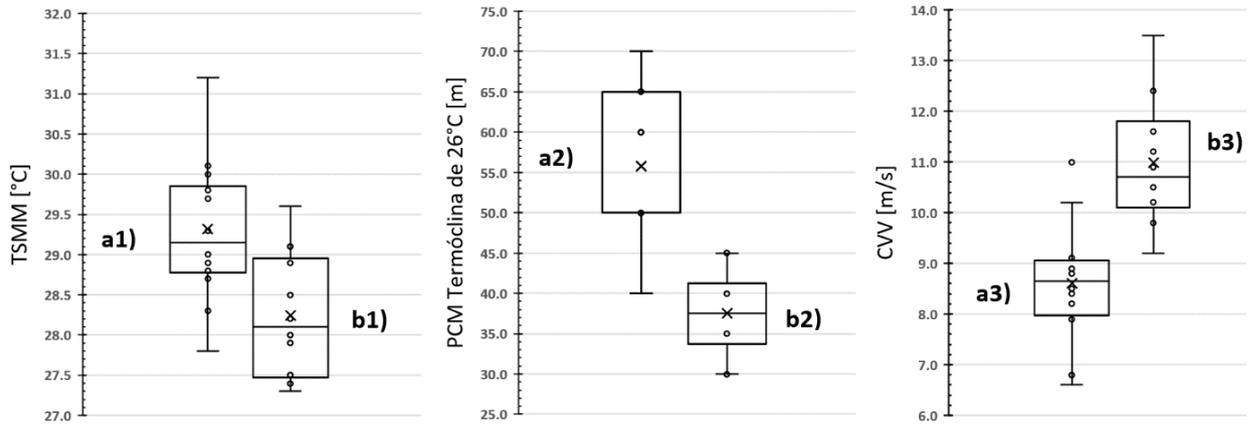

Fuente: Elaboración propia.

## 4. Discusión

En la figura 11, se muestran las gráficas de las anomalías de la TSMM y los valores de la PCM y la cizalladura del viento respecto a la media histórica de los parámetros físicos estudiados en los CTs de RI.

Figura 11. Representación gráfica de la anomalía durante el periodo de RI de la TSMM, PCM para temperaturas mayores de 26,0 °C y magnitud de la cizalladura del viento. La media de cada parámetro se indica con una línea punteada superpuesta en las gráficas.

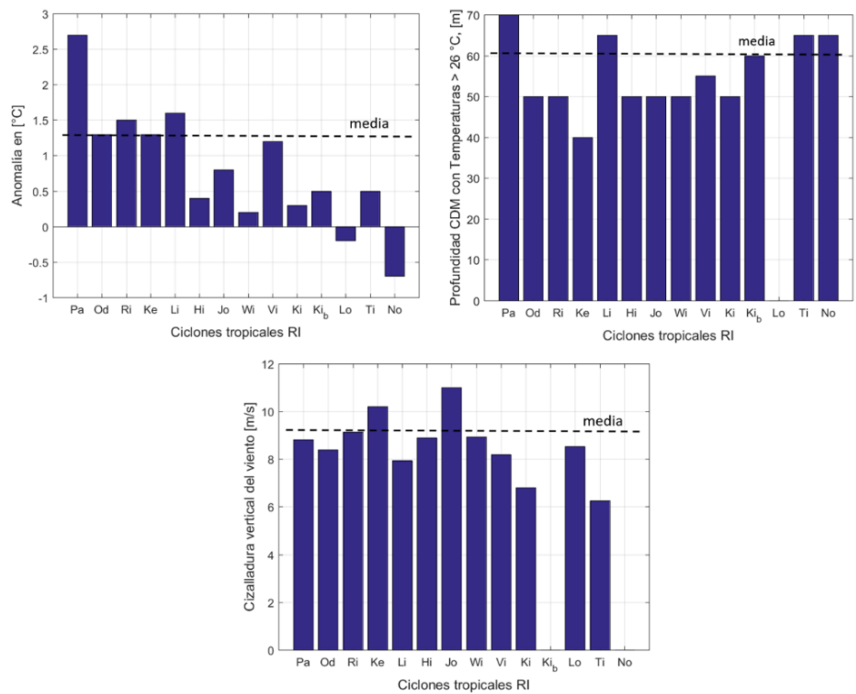

Fuente: Elaboración propia.

Con base en los resultados expuestos en la figura 11, se destacan los valores anómalos de la TSMM en más del 80% de los CTs de RI, particularmente los casos de Patricia, Rick y Linda con anomalías superiores a 1,5 °C. Respecto a la PCM, se observa que, la PCM media en los episodios de RI, fue de 62 m para una temperatura mayor o igual a 26,0 °C. Con base en los datos recabados de la NOAA se reporta una PCM media histórica en los meses de interés de 31 m bajo las mismas condiciones de temperatura lo cual representa una importante anomalía que dota a los CTs de una reserva de energía que permite su intensificación explosiva y prolongada.

En lo que a la cizalladura del viento se refiere, los valores encontrados se encuentran dentro del umbral que permite la intensificación y concuerda con lo dicho en (Montgomery, 2012) «Una baja cizalladura vertical del viento dentro de un ciclón tropical inhibe la transferencia de energía y el proceso de intensificación».

En la figura 12 se muestran las gráficas de las anomalías de la TSMM y los valores de la PCM y la cizalladura del viento respecto a la media histórica de los parámetros físicos estudiados en los CTs no RI.

Figura 12. Representación gráfica de la anomalía de la TSMM, PCM para temperaturas mayores d 26,0 °C y magnitud de la cizalladura del viento de los CTs que no presentaron RI. La media de cada parámetro se indica con una línea punteada superpuesta en las gráficas.

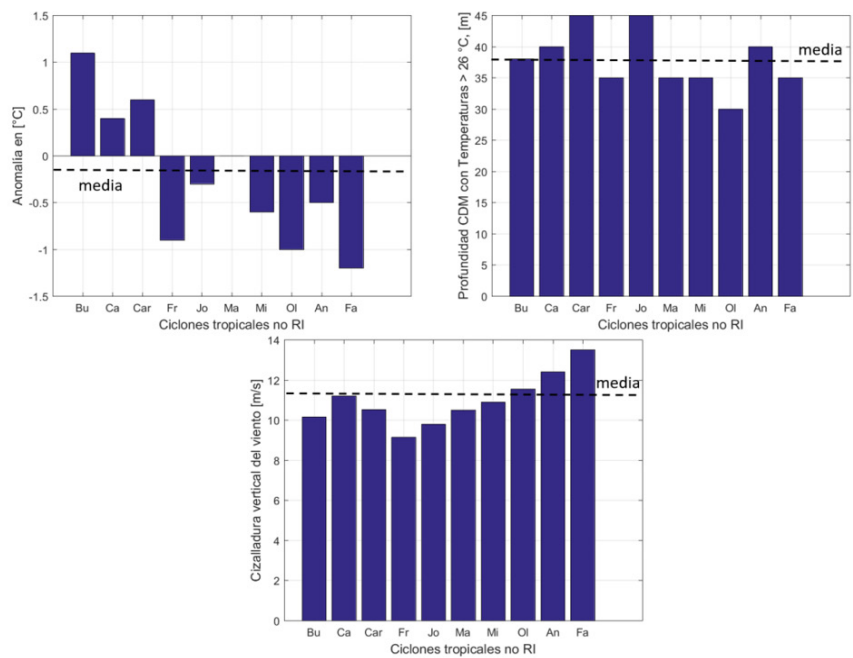

Fuente: Elaboración propia.

Al realizar los cálculos estadísticos de los valores de las tablas 2, 3 y 4 representados de manera gráfica en la figura 12, se observa que variación de la TSMM entre los CTs de RI y los no RI es de 1,2 °C, la variación de la PCM para la termoclina de 26,0 °C es de 30 m y la diferencia de los valores de la cizalladura del viento es de $2,5\ ms^{-1}$.

Con fines comparativos entre los CTs de RI y los que no presentaron esta fase, en la figura 13, se presentan las anomalías porcentuales respecto a la media histórica, de las variables estudiadas en ambos grupos de huracanes. Es importante precisar que cada gráfica es independiente a las otras en cuanto a las anomalías reportadas, ya que el porcentaje de anomalía de un parámetro no necesariamente debe tener interrelación con los otros.

Figura 13. Anomalías porcentuales respecto a la media histórica de las variables física comparadas.

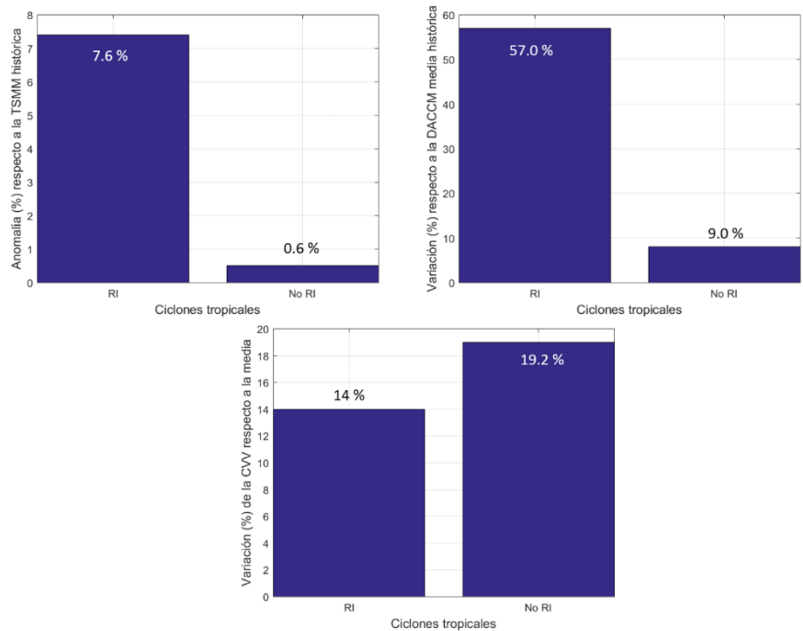

Fuente: Elaboración propia.

Con base en las pruebas de hipótesis realizadas y los resultados que arrojó el algoritmo de clasificación K-means, se comprueba que para los dos grupos de CTs estudiados, los valores umbrales, $\beta_1 = 1,4\ °C, \beta_2 = 30\ m$ y $\beta_3 = -2,5\ ms^{-1}$ estimularon la RI con una confianza del 95%, estos resultados si bien son válidos únicamente para la región de estudio, y bajo las condiciones metodológicas de su recolección y tratamiento, brindan un parteaguas para ser comparados con estudios posteriores en otras regiones e incluso ser cotejados con el trabajo previamente citado de Oropeza y Raga 2015, y verificar si estos valores umbrales, especialmente los térmicos, estuvieron influenciados por remolinos oceánicos anticiclónicos, lo cual brindaría un panorama menos sombrío sobre el pronóstico de intensidad, al menos estadístico, de la probabilidad de que ocurra la RI en los CTs.

## 5. Conclusiones

A modo de conclusión y de manera general, con base en los datos estadísticos obtenidos de los CTs de RI y de la muestra de ciclones que no se intensificó de manera rápida, es evidente la anomalía de los factores térmicos y dinámicos, especialmente la PCM para temperaturas mayores a 26,0 °C presentó la mayor anomalía respecto a la media histórica de 57% que representa más de 30 metros de combustible para la intensificación de los CTs. De la misma manera, la TSMM durante los episodios de RI estaba en promedio 1,4 °C, más cálida que en los registros históricos, es importante destacar el caso del huracán Patricia 2015, que se intensificó con anomalías en la TSMM de más de 2,5 °C y también el hecho de que 12 de los 14 CTs de RI se formaron durante episodios activos del ENOS. Por otro lado, el valor medio de la cizalladura del viento fue de 8,8 $ms^{-1}$, lo cual representa una anomalía de -14% respecto a la media, si bien este valor no es muy bajo, es favorable para que los CTs logren intensificarse, ya que un valor muy bajo o nulo de la cizalladura, inhibe la entrada de aire húmedo, asfixiando la intensificación del sistema tropical.

Con base en la comparativa estadística de ambos grupos de CTs, se encontró que los valores umbrales que favorecieron la RI son, +1,4 °C en la TSMM, +30 m de PCM para la termoclina de 26,0

°C y -2,5 $ms^{-1}$ para el valor de la cizalladura, todos los umbrales respecto a los valores medios de los CTs no RI.

Lo anterior aunado con el hecho que 12 CTs se formaron en los meses de septiembre y octubre refuerza lo dicho por (DeMaría, 2007), existen temporadas en el año en que los factores térmicos dominan sobre los dinámicos, por la gran cantidad de energía que funciona como combustible para el ciclón.

Como era de esperarse, los CTs de RI, son relativamente poco frecuentes en la temporada ciclónica, ya que es necesario un conjunto de factores térmicos y dinámicos y la interacción simultánea de diferentes escalares, dotando al sistema de una energía y estabilidad dinámica estimulante para el desarrollo en intensificación.

Es recomendable realizar estudios similares en otras regiones del océano Pacífico para incrementar el número de datos y realizar construcciones experimentales y numéricas que permitan modelar de mejor manera el mecanismo físico que da lugar a la rápida intensificación de los CTs. Por ejemplo, comparar la muestra de los CTs estudiados en este trabajo con los que investigaron Oropeza y Raga, introduciendo el efecto de los anticiclones oceánicos en la estimulación de la RI.

Otra prueba importante, que puede complementar los resultados obtenidos en este trabajo, es verificar valores umbrales, pero ahora, entre las condiciones medias históricas y las condiciones que estaban presentes durante la RI de los CTs, de la misma manera para los CTs no RI, se esperaría que para los RI los valores de las variables térmicas fueran mayores, pero ¿en qué proporción respecto a los no RI? Y contrastarlos con los obtenidos en esta investigación.

## Contribución de autorías

Mauricio López Reyes: Recopilación de datos, procesamiento de datos, diseño de metodología experimental, análisis de resultados y redacción de al menos 70% del trabajo. Ángel Meulenert Peña: Redacción de 30% (discusiones principalmente) y análisis de los resultados.

## Financiación

## Conflicto de intereses

No hay conflicto de intereses de parte de los autores.

## Bibliografía